# The Backbending Effect in Deformed *e-e* Nuclei

Mohamed E. Kelabi[*]

**Abstract**
An extension of the previously proposed Variable Moment of Inertia with Softness model VMIS has been applied to describe successfully the effect of backbending in some deformed *e-e* nuclei. The model shows good results in the rare earth and actinide regions.

**Introduction**
The effect of backbending[1] has been observed experimentally[2] in the ground state rotational band of some deformed *e-e* nuclei. The effect occurs due to the rapid increase of the moment of inertia with rotational frequency towards the rigid value[3]. When the rotational energy exceeds the energy needed to break a pair of nucleon, the unpaired nucleon goes into different orbits, which result in change of the moment of inertia[4]. An explanation of this effect is due to a disappearance of the pairing correlation[5, 6] by the action of Coriols forces, where the nucleus then undergoes a phase transition from a superfluid state to a state of independent particle motion. Other proposed explanations such as rotational alignment[7, 8, 9] and centrifugal stretching[10], along with the former, could be described in terms of band crossing[11]; the case where the breakup of one pair of nucleons providing a large angular momentum, which may couple with the collective rotation to produce a new band. This effect makes up the backbending phenomena, which is experimentally observed in $\mathcal{J}$ - $\omega^2$ plot as a radical change in the behaviour of the angular momentum from a linearity to a stretching and then back to linear again.

The main purpose of the present work is to investigate, in a phenomenological way, the backbending effect in rare earth and actinide nuclei. It is, hoped to have a good description of the experimentally observed bends by using few parameters formula.

**Theory and Formalism**
From the previously proposed three parametric expression, the Variable Moment of Inertia with Softness model VMIS[12], we have calculated the energy levels in ground state rotational bands (yrast states) of deformed *e-e* nuclei, using the form:

$$E(J) = \frac{A_0}{1+\sigma_1 J} J(J+1) - c\, A_0^3\, \frac{1-2\sigma_1 J}{1+\sigma_1 J}\, J^2(J+1)^2 \qquad (1)$$

where

$$A_0 = \frac{\hbar^2}{2\mathcal{J}_0}$$

---
[*] Physics Department, Al-Fateh University, Tripoli, LIBYA.

and the Softness parameter $\sigma_1$ is given by[13]

$$\sigma_n = \frac{1}{n!} \frac{1}{\mathcal{J}_0} \frac{\partial^n \mathcal{J}(J)}{\partial J^n}\bigg|_{J=0}$$

with $\mathcal{J}_0$ being the unperturbed nuclear moment of inertia[14]; and the constant $c$ is connected with $\beta$- and $\gamma$-vibrational energies through the relation[15]:

$$c = \frac{12}{(\hbar\omega_\beta)^2} + \frac{4}{(\hbar\omega_\gamma)^2}$$

here $\hbar\omega_\beta$ and $\hbar\omega_\gamma$ are the head energies of these vibrations, respectively. However, since the experimental data on $\beta$- and $\gamma$-vibrational bands head energies of deformed doubly even nuclei is incomplete, we take $A_0$, $\sigma_1$, and $c$ as three free parameters of the model, which are adjusted by minimizing Eq. (1) to give a least squares fit to experiment, for low and high angular momenta.

Next we perform a comparative study between our calculations and the experimental data, for the aforementioned nuclei, through the $\mathcal{J} - \omega^2$ plots. The moment of inertia $\mathcal{J}$ and squared rotational frequency $\omega^2$ are related to the spin derivative of the energy[16]

$$\frac{2\mathcal{J}}{\hbar^2} = \left(\frac{d}{d J(J+1)} E(J)\right)^{-1} \qquad (2)$$

$$(\hbar\omega)^2 = \left(\frac{d}{d \sqrt{J(J+1)}} E(J)\right)^2 \qquad (3)$$

respectively, We then employ Eqs. (2) and (3) to deduce the most sensitive relations expressive of $\mathcal{J}$ and $\omega^2$, respectively, giving

$$\frac{2\mathcal{J}}{\hbar^2} = \frac{4J - 2}{\Delta E_\gamma} \qquad (4)$$

$$(\hbar\omega)^2 = (J^2 - J + 1)\left[\frac{\Delta E_\gamma}{2J - 1}\right]^2 \qquad (5)$$

where

$$\Delta E_\gamma = E(J) - E(J-2).$$

**Calculations and Results**
The results of our calculations for the ground state band up to spin $J = 30$, for a set of representative nuclei, along with experimental data[17] are listed in Table 1, and the corresponding parameters $A_0$, $\sigma_1$, and $c$ of Eq. (1) are given in Table 2. The plots of $\mathcal{J} - \omega^2$, that is $\frac{2\mathcal{J}}{\hbar^2}$ versus $(\hbar\omega)^2$ for 22 nuclei covering the mass region from 100 to 242, are shown in Fig. 1 below.

Table 1. Experimental and calculated energy levels in [MeV] of the ground state rotational band of e-e nuclei.

| | | E(2) | E(4) | E(6) | E(8) | E(10) | E(12) | E(14) | E(16) | E(18) | E(20) | E(22) | E(24) | E(26) | E(28) | E(30) |
|---|---|---|---|---|---|---|---|---|---|---|---|---|---|---|---|---|
| 242Pu | Exp | 0.04454 | 0.14730 | 0.30640 | 0.51810 | 0.77860 | 1.08440 | 1.43170 | 1.81670 | 2.23600 | 2.68600 | 3.16300 | 3.66200 | 4.17200 | | |
| | VMIS | 0.04485 | 0.14777 | 0.30623 | 0.51729 | 0.77766 | 1.08373 | 1.43171 | 1.81763 | 2.23742 | 2.68698 | 3.16222 | 3.65912 | 4.17375 | | |
| 240Pu | Exp | 0.04282 | 0.14169 | 0.29432 | 0.49752 | 0.74780 | 1.04180 | 1.37560 | 1.74560 | 2.15200 | 2.59100 | 3.06100 | 3.56000 | 4.08800 | | |
| | VMIS | 0.04378 | 0.14349 | 0.29612 | 0.49857 | 0.74773 | 1.04055 | 1.37405 | 1.74543 | 2.15206 | 2.59152 | 3.06169 | 3.56072 | 4.08709 | | |
| 232Th | Exp | 0.04937 | 0.16212 | 0.33320 | 0.55690 | 0.82700 | 1.13710 | 1.48280 | 1.85860 | 2.26290 | 2.69150 | 3.14420 | 3.61960 | 4.11620 | 4.63180 | 5.16200 |
| | VMIS | 0.04985 | 0.16240 | 0.33262 | 0.55510 | 0.82430 | 1.13479 | 1.48143 | 1.85957 | 2.26517 | 2.69502 | 3.14680 | 3.61927 | 4.11234 | 4.62715 | 5.16625 |
| 230Th | Exp | 0.05320 | 0.17410 | 0.35660 | 0.59410 | 0.87970 | 1.20780 | 1.57290 | 1.97150 | 2.39780 | 2.85000 | 3.32500 | 3.81200 | | | |
| | VMIS | 0.05336 | 0.17372 | 0.35547 | 0.59254 | 0.87872 | 1.20791 | 1.57435 | 1.97286 | 2.39902 | 2.84933 | 3.32141 | 3.81411 | | | |
| 192Os | Exp | 0.20579 | 0.58028 | 1.08923 | 1.70839 | 2.41880 | 3.21100 | | | | | | | | | |
| | VMIS | 0.19671 | 0.58014 | 1.09431 | 1.70956 | 2.41461 | 3.21237 | | | | | | | | | |
| 190Os | Exp | 0.18672 | 0.54785 | 1.05038 | 1.66647 | 2.35700 | | | | | | | | | | |
| | VMIS | 0.18165 | 0.54872 | 1.05358 | 1.66396 | 2.35753 | | | | | | | | | | |
| 182Os | Exp | 0.12700 | 0.40040 | 0.79380 | 1.27780 | 1.81210 | 2.34630 | 2.84090 | 3.32020 | | | | | | | |
| | VMIS | 0.18165 | 0.40327 | 0.79952 | 1.27937 | 1.80503 | 2.34007 | 2.85229 | 3.31633 | | | | | | | |
| 180Os | Exp | 0.13211 | 0.40862 | 0.79508 | 1.25744 | 1.76757 | 2.30871 | 2.87500 | | | | | | | | |
| | VMIS | 0.12954 | 0.40637 | 0.79471 | 1.25913 | 1.77043 | 2.31055 | 2.87660 | | | | | | | | |
| 172Os | Exp | 0.22777 | 0.60617 | 1.05447 | 1.52495 | 2.02387 | 2.56450 | 3.19940 | 3.82330 | 4.51070 | | | | | | |
| | VMIS | 0.23150 | 0.60350 | 1.04260 | 1.52323 | 2.03755 | 2.58635 | 3.17572 | 3.81552 | 4.51873 | | | | | | |
| 170Os | Exp | 0.28670 | 0.74990 | 1.32542 | 1.94580 | 2.54520 | | | | | | | | | | |
| | VMIS | 0.23150 | 0.60350 | 1.04260 | 1.52323 | 2.03755 | | | | | | | | | | |
| 162Er | Exp | 0.10204 | 0.32961 | 0.66664 | 1.09668 | 1.60282 | 2.16511 | 2.74572 | 3.29240 | 3.84660 | | | | | | |
| | VMIS | 0.10072 | 0.32781 | 0.66725 | 1.10073 | 1.60651 | 2.16024 | 2.73572 | 3.30568 | 3.84241 | | | | | | |

Table 1 (continue). Experimental and calculated energy levels in [MeV] of the ground state rotational band of e-e nuclei.

| | E(2) | E(4) | E(6) | E(8) | E(10) | E(12) | E(14) | E(16) | E(18) | E(20) | E(22) | E(24) | E(26) | E(28) | E(30) |
|---|---|---|---|---|---|---|---|---|---|---|---|---|---|---|---|
| 160Dy Exp | 0.08679 | 0.28382 | 0.58118 | 0.96685 | 1.42872 | 1.95150 | 2.51500 | 3.09170 | 3.67220 | | | | | | |
| VMIS | 0.08699 | 0.28449 | 0.58307 | 0.97046 | 1.43208 | 1.95150 | 2.51086 | 3.09131 | 3.67341 | | | | | | |
| 152Nd Exp | 0.07251 | 0.23662 | 0.48395 | 0.80540 | 1.19530 | 1.64760 | 2.15790 | 2.72240 | 3.33570 | | | | | | |
| VMIS | 0.07344 | 0.23742 | 0.48377 | 0.80517 | 1.19507 | 1.64773 | 2.15813 | 2.72202 | 3.33588 | | | | | | |
| 150Nd Exp | 0.13521 | 0.38145 | 0.72040 | 1.12970 | 1.59900 | 2.11900 | 2.68250 | | | | | | | | |
| VMIS | 0.12766 | 0.38034 | 0.72251 | 1.13269 | 1.59884 | 2.11579 | 2.68381 | | | | | | | | |
| 140Ba Exp | 0.60235 | 1.13059 | 1.66070 | 2.46890 | 3.38360 | | | | | | | | | | |
| VMIS | 0.12766 | 1.12097 | 1.70398 | 2.42745 | 3.39533 | | | | | | | | | | |
| 132Nd Exp | 0.21262 | 0.60980 | 1.13120 | 1.71040 | 2.30910 | 2.94480 | 3.63010 | 4.36900 | 5.17990 | 6.06250 | 7.00630 | 8.00780 | | | |
| VMIS | 0.22679 | 0.62678 | 1.12526 | 1.68772 | 2.29814 | 2.95055 | 3.64522 | 4.38679 | 5.18332 | 6.04564 | 6.98710 | 8.02332 | | | |
| 130Nd Exp | 0.15905 | 0.48550 | 0.93994 | 1.48710 | 2.10041 | 2.76390 | 3.46810 | 4.21160 | 5.02090 | 5.91830 | 6.90920 | 7.99390 | | | |
| VMIS | 0.17278 | 0.50900 | 0.95814 | 1.49023 | 2.08785 | 2.74204 | 3.44991 | 4.21331 | 5.03796 | 5.93291 | 6.91020 | 7.98460 | | | |
| 122Xe Exp | 0.33118 | 0.82830 | 1.46657 | 2.21730 | 3.03950 | 3.91920 | 4.90000 | | | | | | | | |
| VMIS | 0.30400 | 0.83086 | 1.48237 | 2.21925 | 3.02991 | 3.91887 | 4.90178 | | | | | | | | |
| 120Ba Exp | 0.18600 | 0.54400 | 1.04020 | 1.64500 | 2.33600 | 3.08300 | | | | | | | | | |
| VMIS | 0.30400 | 0.83086 | 1.04305 | 1.64701 | 2.33267 | 3.08397 | | | | | | | | | |
| 110Pd Exp | 0.37380 | 0.92077 | 1.57404 | 2.29600 | 3.13100 | 4.03000 | | | | | | | | | |
| VMIS | 0.17986 | 0.54328 | 1.57819 | 2.30588 | 3.11566 | 4.03507 | | | | | | | | | |
| 102Zr Exp | 0.15178 | 0.47828 | 0.96478 | 1.59490 | 2.35150 | 3.21230 | | | | | | | | | |
| VMIS | 0.36246 | 0.92307 | 0.96732 | 1.59754 | 2.35086 | 3.21270 | | | | | | | | | |
| 100Zr Exp | 0.21253 | 0.56449 | 1.05140 | 1.68740 | 2.42550 | 3.27210 | 4.20850 | | | | | | | | |
| VMIS | 0.18290 | 0.55415 | 1.06745 | 1.69521 | 2.42548 | 3.26004 | 4.21327 | | | | | | | | |

Table 2. Fitted parameters of Eq. (1).

| Nucl. | $A_0$ [MeV] | $\sigma_1$ | $c$ [MeV]$^{-2}$ |
|---|---|---|---|
| 242Pu | 0.00755 ± 0.00002 | 0.00437 ± 0.00021 | 3.98078 ± 0.04013 |
| 240Pu | 0.00741 ± 0.00002 | 0.00766 ± 0.00029 | 2.50813 ± 0.10532 |
| 232Th | 0.00848 ± 0.00002 | 0.00915 ± 0.00013 | 5.47149 ± 0.05142 |
| 230Th | 0.00908 ± 0.00003 | 0.00926 ± 0.00023 | 5.25314 ± 0.03243 |
| 192Os | 0.03769 ± 0.00090 | 0.07411 ± 0.00527 | 0.18735 ± 0.03196 |
| 190Os | 0.03377 ± 0.00070 | 0.05762 ± 0.00383 | 0.02066 ± 0.21415 |
| 182Os | 0.02179 ± 0.00026 | 0.01067 ± 0.00087 | 4.05569 ± 0.07013 |
| 180Os | 0.02264 ± 0.00012 | 0.01748 ± 0.00039 | 4.49207 ± 0.05651 |
| 172Os | 0.05356 ± 0.00252 | 0.19396 ± 0.01449 | 0.01841 ± 0.00173 |
| 170Os | 0.06002 ± 0.00211 | 0.14741 ± 0.01143 | -0.05931 ± 0.02665 |
| 162Er | 0.01706 ± 0.00021 | 0.00526 ± 0.00109 | 3.45923 ± 0.03165 |
| 160Dy | 0.01470 ± 0.00008 | 0.00467 ± 0.00057 | 3.38056 ± 0.01735 |
| 152Nd | 0.01263 ± 0.00068 | 0.01564 ± 0.01539 | 0.42559 ± 6.22898 |
| 150Nd | 0.02414 ± 0.00044 | 0.06695 ± 0.00354 | 0.23934 ± 0.04818 |
| 140Ba | 0.53745 ± 0.34584 | 2.20139 ± 1.56879 | 0.00024 ± 0.00044 |
| 132Nd | 0.04763 ± 0.00125 | 0.12998 ± 0.00568 | 0.02195 ± 0.00129 |
| 130Nd | 0.03315 ± 0.00076 | 0.07548 ± 0.00365 | 0.07426 ± 0.00378 |
| 122Xe | 0.06500 ± 0.00278 | 0.14129 ± 0.01198 | 0.02645 ± 0.00305 |
| 120Ba | 0.03344 ± 0.00050 | 0.05772 ± 0.00273 | 0.00962 ± 0.07407 |
| 110Pd | 0.08781 ± 0.00412 | 0.22669 ± 0.01849 | 0.01796 ± 0.00167 |
| 102Zr | 0.02557 ± 0.00039 | 0.01548 ± 0.00371 | 0.70338 ± 0.26768 |
| 100Zr | 0.03383 ± 0.00180 | 0.05395 ± 0.00981 | 0.33311 ± 0.06559 |

Fig. 1. Plots of $\mathcal{J} - \omega^2$.

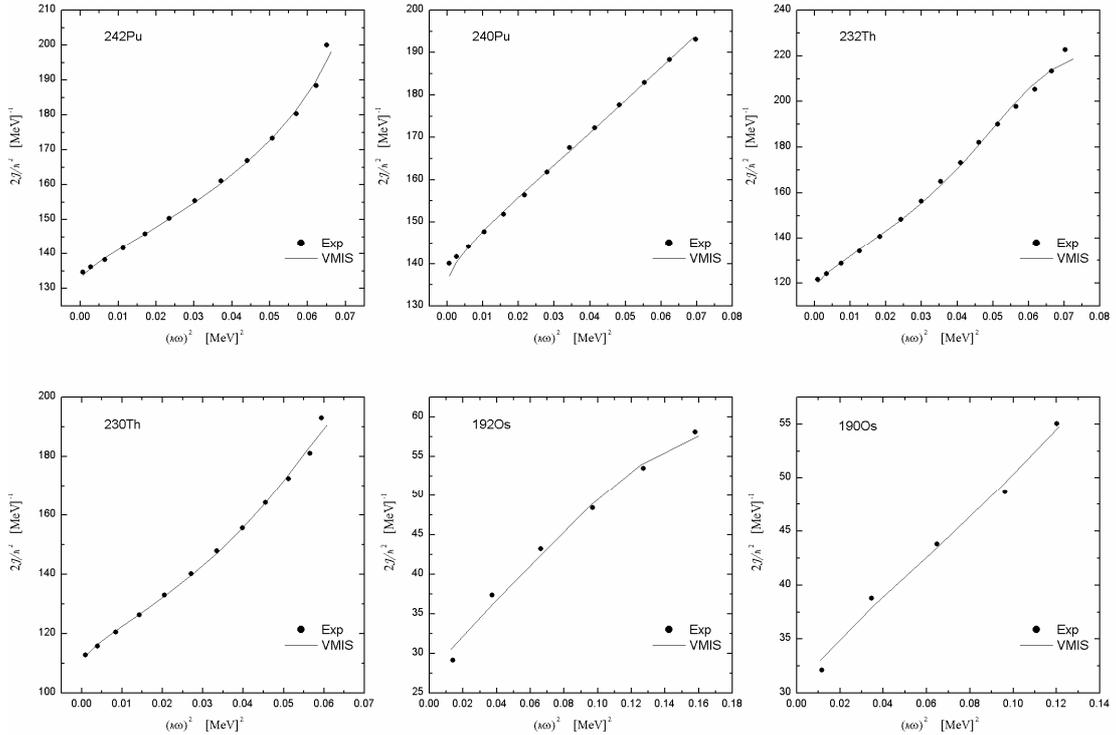

Fig. 1. (continue) Plots of $\mathcal{J}$ - $\omega^2$.

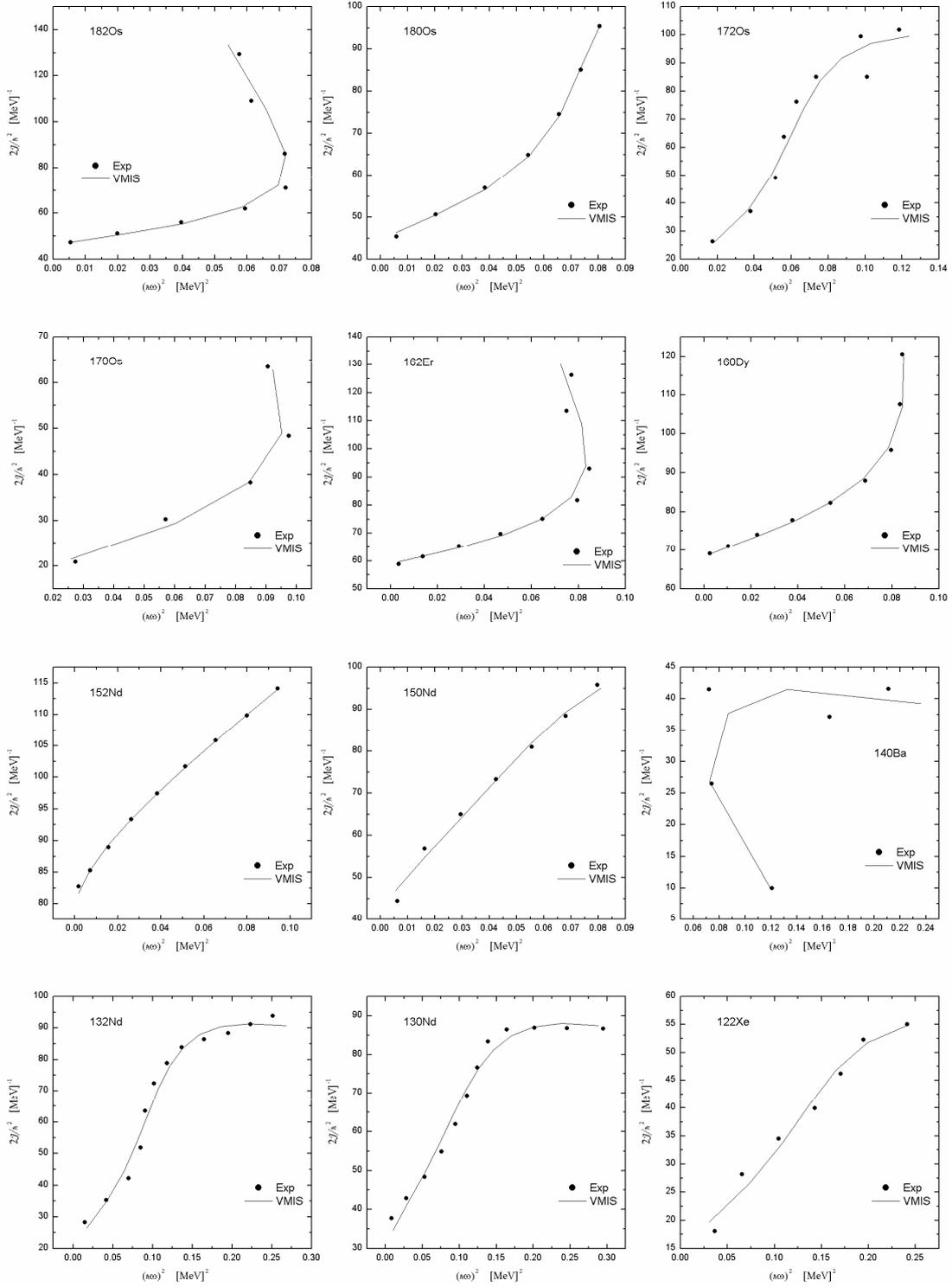

Fig. 1. (continue) Plots of $\mathcal{J}$ - $\omega^2$.

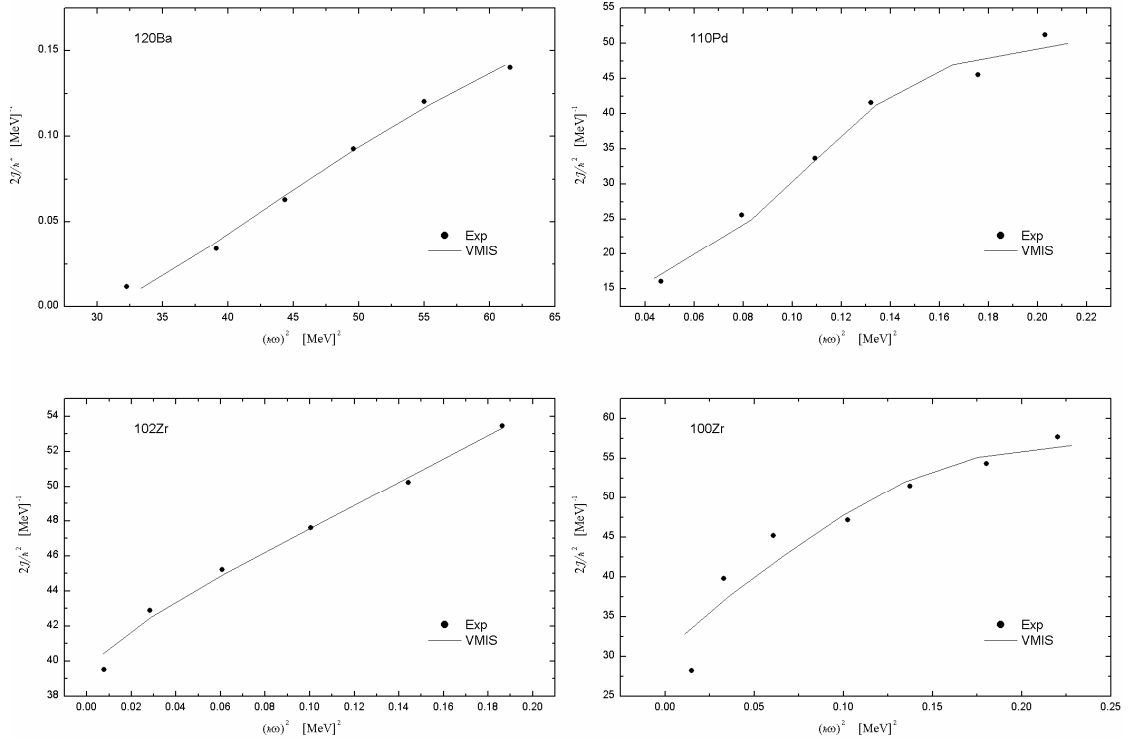

**Conclusion**

The present work suggests that the Variable Moment of Inertia with Softness model VMIS is a successful tool in studying ground state energy levels in deformed nuclei up to high spin states. Predictions of the model give fairly accurate description of the backbending phenomena in rare earth and actinide regions. The appearance of backbending can be attributed due to the smallness of critical rotational frequency of the $\pi - i_{13/2}$ protons than for $\nu - j_{15/2}$ neutrons, therefore, the behaviour of the proton pair $i_{13/2}$ at high spins seems to be decisive for such effect. On the other hand, the absence of backbending in some nuclei may be ascribed to the presence of stable octupole deformation in them.